%% file: Template.tex
\documentclass{article}
\usepackage{spconf,amsmath,graphicx}
\usepackage[dvipsnames]{xcolor}
\usepackage[utf8]{inputenc}
\usepackage[english]{babel}
\usepackage{float}
\usepackage{graphicx}
\usepackage{multirow}
\usepackage{bm}
\usepackage[section]{placeins}
\usepackage[normalem]{ulem}
\useunder{\uline}{\ul}{}
\usepackage{url}
\usepackage[breaklinks]{hyperref}   
           
\usepackage{caption}
\usepackage{listings}
\usepackage{algorithm}
\usepackage{algorithmic}
\usepackage{amsmath,amssymb}
\usepackage{graphicx}
\usepackage{url}
\usepackage{multirow}
\usepackage{bm}
\usepackage{mathrsfs,amsmath}
\usepackage[section]{placeins}
\usepackage[normalem]{ulem}
\useunder{\uline}{\ul}{}
\usepackage{caption} 
\usepackage{stfloats}
\usepackage{siunitx}

\usepackage{booktabs}
\usepackage{array} % <-- new
\newcolumntype{C}[1]{>{\centering\arraybackslash}p{#1}} % <-- new

\usepackage[caption=false,font=normalsize,labelfont=sf,textfont=sf]{subfig}
\usepackage{enumitem} % in your preamble

\usepackage[caption=false,font=normalsize,labelfont=sf,textfont=sf]{subfig}

\title{Vision-Inspired Image Quality Assessment for Radar-Based Human Activity Representations}
\name{Huy Trinh, Davis Liu, Munia Humaira, Peter Lee, Zhou Wang}
\address{University of Waterloo}

\begin{document}
%\ninept
%
\maketitle
\begin{abstract}
\label{sec:abs}
\input{tex/abstract}

\end{abstract}
\section{Introduction}
\label{sec:intro}
\input{tex/introduction}
\section{DYNAMIC ACTIVITIES EVALUATION}
\label{sec:dynamic_activities_evaluation}
\input{tex/dynamic}
\section{PROPOSED PRE-PROCESSING METHODS FOR STATIC ACTIVITIES' FEATURE MAPS}
\label{sec:proposed_methods}
\input{tex/proposed_method}
\section{CONCLUSION}
\input{tex/conclusion}

% \vfill\pagebreak

\bibliographystyle{IEEEbib}
\bibliography{strings,refs}

\end{document}

%% file: tex/abstract.tex
Radar-based human activity recognition has gained attention as a privacy-preserving alternative to vision and wearable sensors, especially in sensitive environments like long-term care facilities. Micro-Doppler spectrograms derived from FMCW radar signals are central to recognizing dynamic activities, but their effectiveness is limited by noise and clutter. In this work, we use a benchmark radar dataset to reimplement and assess three recent denoising and preprocessing techniques: adaptive preprocessing, adaptive thresholding, and entropy-based denoising. To illustrate the shortcomings of conventional metrics in low-SNR regimes, we evaluate performance using both perceptual image quality measures and standard error-based metrics. We additionally propose a novel framework for static activity recognition using range-angle feature maps to expand HAR beyond dynamic activities. We present two important contributions: a temporal tracking algorithm to enforce consistency and a no-reference quality scoring algorithm to assess RA-map fidelity. According to experimental findings, our suggested techniques enhance classification performance and interpretability for both dynamic and static activities, opening the door for more reliable radar-based HAR systems.

%% file: tex/introduction.tex
% \textcolor{red}{Wwrite this paragraph shorter as not a review paper}\\
% \textcolor{red}{Emphasize the importance of treat spectrograms and heatmaps as image-centric for machine learning and deep learning applications}\\
% Radar technology is quite mature in many sectors, such as aerospace, military, and automotive. An emerging field of study is radar-based human activity recognition. In certain settings like long-term care resident monitoring, human activity recognition must be non-invasive and maintain a high level of privacy. Despite satisfying performance requirements, RGB cameras and wearable devices have raised concerns about security and privacy; radar presents a potential solution. However, raw radar signals are often contaminated with noise, affecting the reliability and quality of the extracted information. Noise comes from various sources, such as thermal noise from sensors, electronic interference, clutter from the surrounding environment, etc. It can raise the background floor and obscure weak target reflections. In addition, many ongoing investigations have used very low-resolution radars, which are susceptible to noise contamination when extracting signals.

Radar technology possesses a longstanding history of application across various sectors, including aerospace, defence, and automotive systems. More recently, especially in settings where non-invasiveness and privacy are crucial, such as long-term care (LTC) facilities, it has surfaced as a promising tool for human activity recognition (HAR). Wearable devices and RGB cameras sometimes generate privacy and security issues despite their performance, but radar provides a contactless and privacy-preserving option. Noise in the raw radar signals from sources, including thermal noise, electrical interference, and environmental clutter, challenges the dependability of radar-based HAR. These elements can hide weak reflections and raise the background floor, therefore hindering precise activity detection. Moreover, many present studies depend on low-resolution radar systems, which are susceptible to noise contamination when extracting signals \cite{10784889}, \cite{10880536}.

Extensive research has been done on various preprocessing and denoising methods that focus on dynamic and basic activity recognition tasks. Deep learning-driven techniques, such as inferencing and testing on machine learning models, are then used to assess how effective these algorithms are. However, their numerical impact, visual explainability of models, and method comparisons have not been thoroughly examined. Additionally, to the best of our knowledge, there aren't many studies on the recognition of static activities or slowly changing motions. The main contributions of our research can be summarized in two points:

% There have been extensive research projects on various preprocessing and denoising methods, focusing on dynamics and basic activity recognition tasks. The effectiveness of these algorithms is then evaluated using deep learning-driven methods (e.g., inferencing and testing on machine learning models). However, their visual explainability of models, numerical impact, and comparison between methods have not been investigated and explored. In addition, to our best of knowledge, there have not been many research works on slow-changing motions or static activities recognition. Our contributions in this study are summarized as follows:
\begin{itemize}[leftmargin=*, noitemsep, topsep=0pt]
    \item We quantitatively evaluate the efficiency of recent radar-spectrogram-based 
    denoising and preprocessing methods using error-based metrics and image quality assessment.
    \item We propose algorithms for scoring quality and tracking Range–Angle (RA) feature maps for stationary activities.
\end{itemize}

% \textcolor{red}{[NOT SURE IF WE NEED TO POINT OUT THE LIMIT OF ALI's PAPER ABOUT USING ML TO CLASSIFY ACTIVITIES BASED ON POSITION RATHER ACTUAL LEARNING ~ Try to understand Multi-dimensional feature maps, its advantage,]}

% \textcolor{red}{Also point out not much work study static activities which is more alike to elderly's in LTC, the challenges of FMCW Doppler radar to detect these,}
% Radar spectrogram-based heatmaps such as Range-Doppler and Doppler-time are widely used in training deep learning models in various recognition tasks. However, there has been a lack of quantitative evaluation of these algorithms. 

% \textcolor{blue}{[Human Detection from 4D Radar Data in Low-Visibility Field Conditions}

%% file: tex/dynamic.tex
% \textcolor{orange}{Showcase some, emphasize the lack of quantiative evaluation of denoising A lot of denoising methods (we can take some latest noticeable works from our review report), but most of them are quantified by deep learning driven approaches to propose that the model can utilize these}

% \textcolor{orange}{Demonstrate 2-3 denoising papers that evaluate their algorithm} 
% \textcolor{red}{Munia: Write about dynamic activity recognition}

Since radar spectrograms containing human activities are often very noisy, it is important to investigate multiple denoising approaches to obtain the optimum one for our use case. To achieve this, we replicated and reimplemented the three latest denoising and preprocessing methods in the radar-based HAR domain.

\subsection{Adaptive Preprocessing}
% Park et al. \cite{park2024resolution} suggest an adaptive resolution preprocessing (APr) method to improve micro-Doppler (mD) spectrogram-based HAR using frequency-modulated continuous wave (FMCW) radar signals. The full detail of a variety of human activities is frequently not captured by traditional spectrograms, which are based on fixed time–frequency resolution. \textcolor{red}{Is this duplicate with above? The authors present a resolution-adaptive spectrogram, which nonlinearly modifies the frequency resolution according to the energy distribution of the mD signatures to get around this restriction.} To improve resolution in the lower-frequency region, which is usually where subtle human motion features are concentrated, the suggested method first determines the active frequency range in a given spectrogram before applying a logarithmic mean-square scaling function. As a result, the spectrogram representation for HAR has better discriminative power. Deep learning models trained on RA spectrograms consistently perform better than those trained on conventional representations, according to experimental results.
An adaptive resolution preprocessing (APr) technique has been suggested by Park et al. \cite{park2024resolution} to improve micro-Doppler (mD) spectrogram-based HAR using frequency-modulated continuous wave (FMCW) radar signals. Conventional spectrograms employ a fixed time-frequency resolution, which frequently falls short of capturing the characteristics of the entire range of motion, particularly for low-frequency or subtle movements. The authors address this by introducing a resolution-adaptive spectrogram, which uses the energy distribution of the mD signatures to non-linearly adjust frequency resolution. In particular, the technique locates each spectrogram's active frequency range and uses logarithmic mean-square scaling to improve resolution in lower-frequency bands, which are usually where fine motion details show up. Compared to traditional representations, this adaptive method generates more discriminative spectrograms, which enhances the performance of the deep learning model.

\subsection{Adaptive-Thresholding}
An adaptive thresholding algorithm (ATh) proposed by Li et al.\cite{Li2023RadarbasedHA} offers a low-computation, high-accuracy solution for radar-based human activity recognition, making it well-suited for multi-domain mD signatures and resource-constrained systems. The method consists of three stages: signal preprocessing, adaptive thresholding, and classification. In preprocessing, a Hamming-windowed Fast Fourier Transform (FFT) generates a range-time map from time-domain radar pulses, followed by a $4^\text{th}$ order Butterworth filter\cite{Butterworth1930} to suppress noise. A Short-Time Fourier Transform (STFT), using a $0.2$ second Hamming window with $95\%$ overlap, produces the mD spectrogram. Adaptive thresholding converts the spectrogram to grayscale and iteratively updates a threshold based on average pixel intensity until convergence (typically at $V=0.1$). The resulting binary mask is used to generate masked spectrograms and phase-related images. The final classification is performed using a support vector machine (SVM) algorithm.

\subsection{Entropy-Based Denoising}
The third reimplemented algorithm, proposed by Nguyen et al.~\cite{entropy_nguyen}, applies a two-step entropy-based denoising (EBD) approach for spectrogram enhancement and classification. The first step identifies the optimal range-bin interval containing the most motion-relevant signal energy. This is achieved by computing the FFT of each range-bin across the slow-time axis, identifying the most frequently occurring peak location, and evaluating surrounding intervals via STFT. The interval with the lowest average Shannon entropy, $H_{avg}$, is selected as optimal.
\begin{equation} r_{opt} = \arg \min_{1 \leq q \leq Q} H_{avg}(range_{r_q}) 
\label{eq 1.} 
\end{equation} 
In the second step, adaptive spectrogram thresholding is applied within the selected interval. A Gaussian-windowed STFT is computed frame-by-frame, and a dynamic cut-threshold is generated based on local average energy. This masking function reduces background noise while preserving key micro-Doppler signatures, ultimately improving the clarity of the spectrogram for downstream classification tasks.

\subsection{Experiment Setup}
\begin{figure}[htbp]
\centerline{\includegraphics[width=\linewidth]{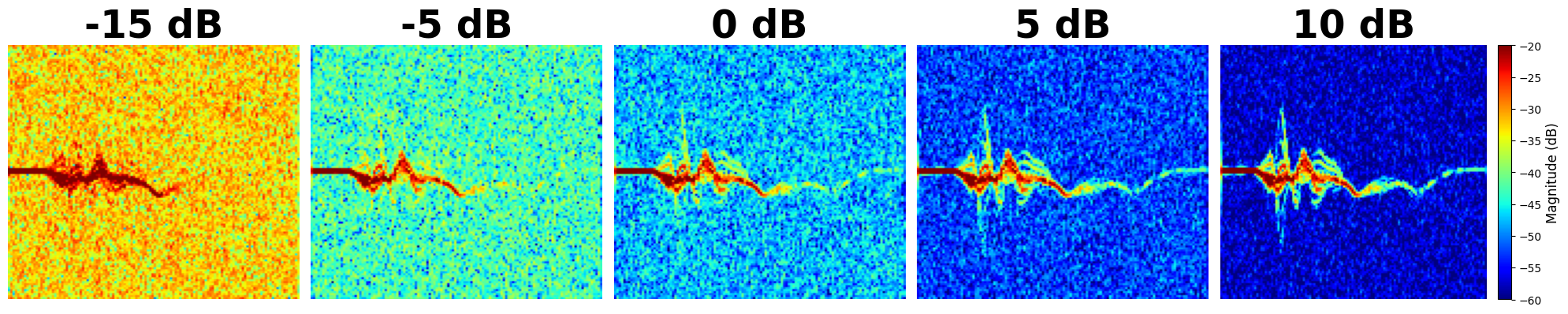}}
\caption{Spectrogram after adding 5 different White Gaussian noise levels.}
\label{fig:all_noise_levels}
\end{figure}

After implementing the algorithms discussed above, we evaluated their performance using the dataset provided in \cite{entropy_nguyen}. The raw signal dataset comprises $19,800$ samples, generated by combining $11$ distinct activities, $5$ aspect angles, and $60$ iterations. The first step in our processing pipeline involves generating spectrograms for each radar signal file. Following this, white Gaussian noise (WGN) is added at $5$ different signal-to-noise ratio (SNR) levels: $-10$\ dB, $-5$\ dB, $0$\ dB, and $10$\ dB. The resulting spectrograms are illustrated in Figure \ref{fig:all_noise_levels}. These noisy spectrograms are then processed using three reimplemented denoising methods, as well as our combined APr and ATh method (APr+ATh).
% Uncomment this for full page width figure if we still have space
% \begin{figure*}[!t]
%   \centering
%   \includegraphics[width=\textwidth]{figures/all_noise_levels.png}
%   \caption{Spectrograms of walking under five different white Gaussian noise levels. Spectrogram after adding 5 different White Gaussian noise levels  Note how the micro‑Doppler signatures become increasingly obscured at lower SNRs.}
%   \label{fig:all_noise_levels}
% \end{figure*}
The Structural Similarity Index Measure (SSIM) \cite{Zhou_SSIM}, a perceptual metric that evaluates image quality based on structural fidelity, is used to evaluate the denoised spectrogram outputs, which have been done quite rarely before \cite{s20175007}. The similarity between image structure and contrast is the main focus of SSIM. We used several common quantitative evaluation metrics, such as Mean Squared Error (MSE), Mean Absolute Error (MAE), Root Mean Squared Error (RMSE), Peak Signal-to-Noise Ratio (PSNR), and the Pearson Correlation Coefficient.

\begin{figure}[htbp]
\centerline{\includegraphics[width=\linewidth]{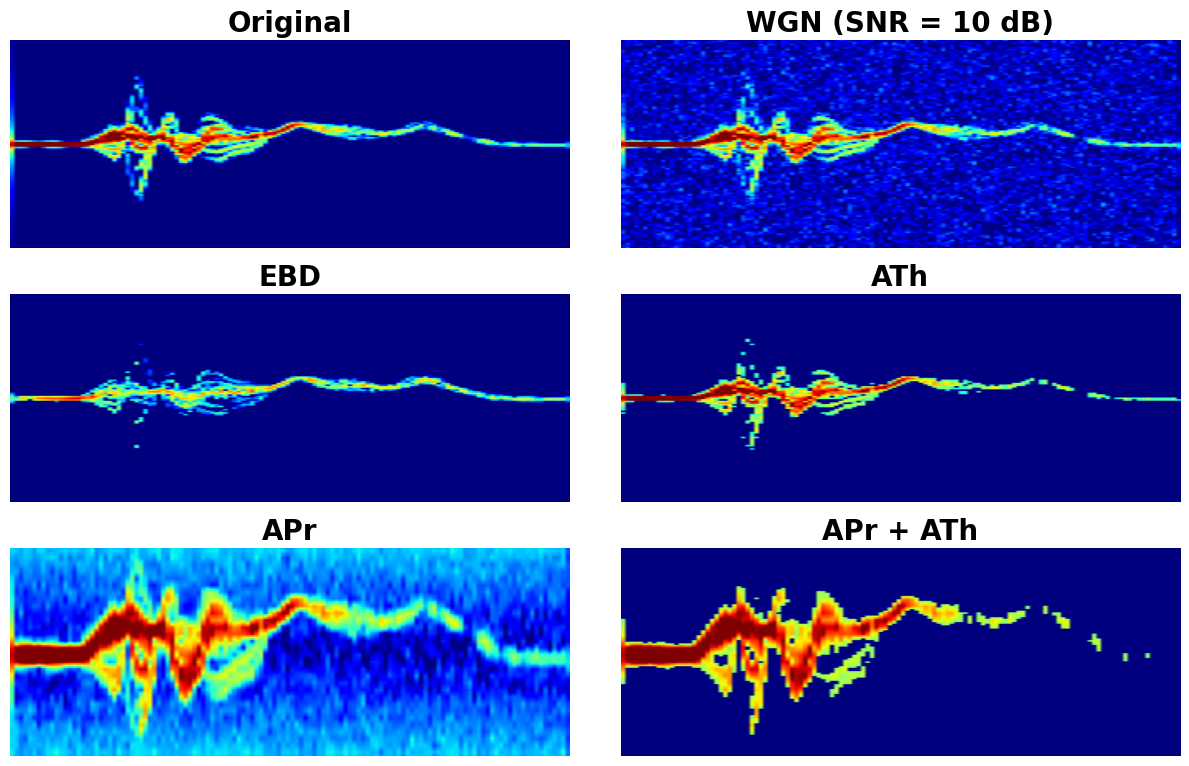}}
\caption{Spectrogram results for Walking (W): original spectrogram (top-left), noisy spectrogram (top-right) and 4 methods under WGN (SNR = 10 dB).}
\label{fig:all_figs_snr_10}
\end{figure}

Figure \ref{fig:all_figs_snr_10} shows the result of $4$ methods along with the original noise-free and noisy spectrograms. Its corresponding evaluation metrics are shown in Table \ref{tab:table_fig1}. It can be inferred from the table that the ATh algorithm removes noise selectively and preserves most of the detail in case SNR = $5$\ dB. EBD algorithm, however, preserves large-scale intensity distribution with high correlation but degrades to local structures. The APr algorithm effectively adjusted the frequency resolution, enabling mD signatures to span a broader frequency range. This ensures optimal representation by dynamically expanding the active frequency, though it also has the unintended effect of scaling noise within the calculated range. Applying ATh after APr removes these noises for a cleaner representation but can also remove fine main signal details in this case.

\begin{table}[H]
\centering
\caption{Evaluation metrics result for walking (W) under White Gaussian Noise (SNR = 10 dB).}
% \textbf{Bold} indicates the best (optimal) value in each row.}
\scalebox{0.8}{
\begin{tabular}{lccccc}
\toprule
\textbf{METRICS} & \textbf{NS} & \textbf{EBD} & \textbf{ATh} & \textbf{APr} & \textbf{APr + ATh} \\ 
\midrule
MSE           & 5053.2   & 12953.8  & \textbf{4785.7} & 6254.1 & 5043.3\\
MAE           & 63.82     & 107.48    & \textbf{62.15} & 73.05 & 65.58\\
RMSE          & 71.08    & 113.9     &\textbf{69.17}    &79.08 &71.01\\
PSNR (dB)     & -19.69    & -23.78   & \textbf{-19.46} &-20.62 &-19.68\\
Pearson Corr & 0.572      & \textbf{0.6168}     & 0.6006 &0.4462 & 0.5561\\
SSIM          & 0.365      & 0.3639    & \textbf{0.4109} &0.2881 & 0.3227  \\
\bottomrule
\multicolumn{6}{l}{\footnotesize\textbf{Bold} indicates the best (optimal) value in each row}
\end{tabular}
}
\label{tab:table_fig1}
\end{table}

\begin{figure}[htbp]
\centerline{\includegraphics[width=\linewidth]{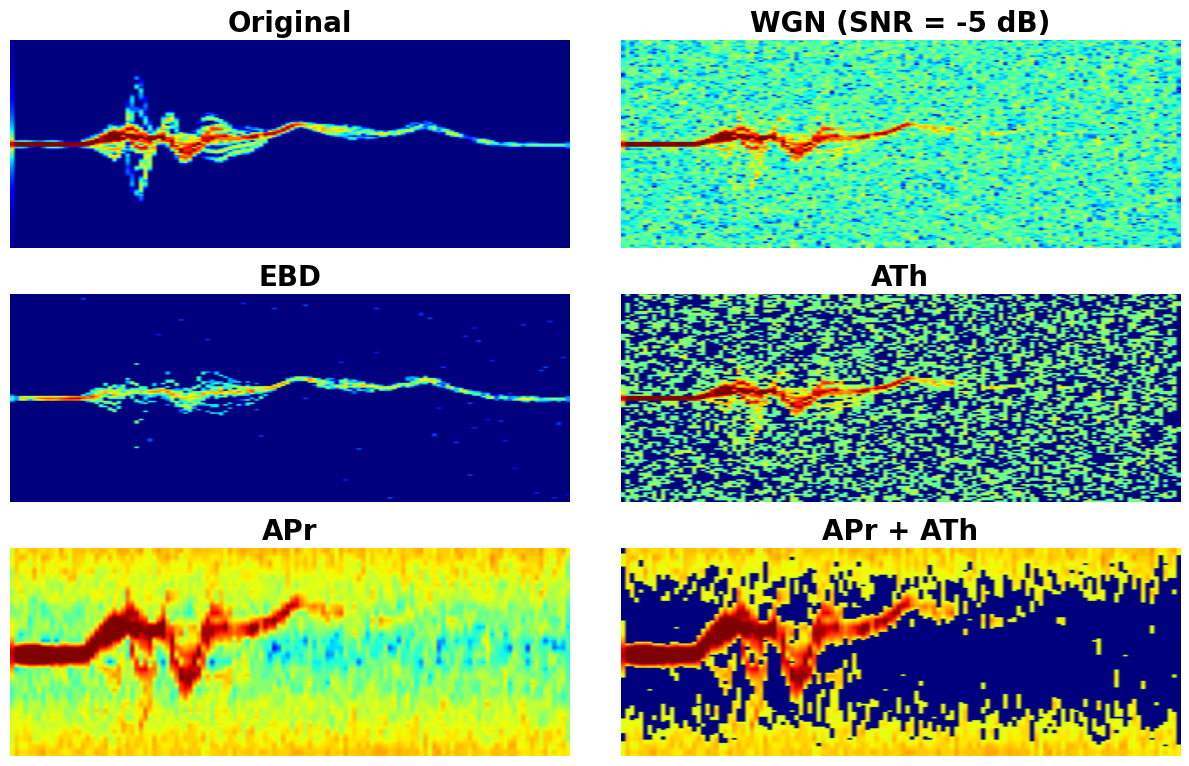}}
\caption{Spectrogram results for Walking (W): original spectrogram (top-left), noisy spectrogram (top-right) and 4 methods under WGN (SNR = -5 dB).}
\label{all_figs_snr_-5}
\end{figure}

\begin{table}[H]
\centering
\caption{Evaluation metrics result for walking (W) under White Gaussian Noise (SNR = -5 dB).}
\scalebox{0.8}{
\begin{tabular}{lccccc}
\toprule
\textbf{METRICS} & \textbf{NS} & \textbf{EBD} & \textbf{ATh} &\textbf{APr} & \textbf{APr + ATh}\\ 
\midrule
MSE           & 7184.1  & 13501.9  & \textbf{6152.6} &8365.6 & 6935.6   \\
MAE           & 77.81    & 110.18    & \textbf{71.01} &84.86 &75.32 \\
RMSE          &84.76    &116.19 &\textbf{78.43} &91.46 &83.28\\
PSNR (dB)     & -21.22    & -23.96    & \textbf{-20.55} &-21.8864 & -21.07 \\
Pearson Corr & 0.329      & \textbf{0.533}     & 0.225 & 0.073 & -0.058 \\
SSIM          &  0.261      &  \textbf{0.305}    & 0.175  &0.227  &0.177\\
\bottomrule
\multicolumn{6}{l}
{\footnotesize\textbf{Bold} indicates the best (optimal) value in each row}
\end{tabular}
}
\label{tab:evaluation_results}
\end{table}

Under severe noisy (–5 dB) condition (Figure \ref{all_figs_snr_-5}, Table \ref{tab:evaluation_results}), EBD shows a better ability to denoise although there are some speckles left in the background, which may explain lower metrics compared to its corresponding values in Table \ref{tab:table_fig1}. ATh is quite prone to very high noise levels, whose SNRs are negative. However, its pixel error (MSE, RMSE, PSNR) metrics are lower than those of EBD. SSIM, on the other hand, shows a more reasonable alignment in both cases. Thus, SSIM can be used as a cost function when training deep learning models for fidelity-based applications such as Generative Adversarial Network (GAN)-based \cite{synthesis_GAN} reconstruction or synthetic spectrogram generators. 

According to our study, depending only on error-based metrics can be deceptive, especially when there is a low SNR as perceptual fidelity rather than pixel-level similarity is needed to preserve subtle mD signatures, which are essential for precise human activity recognition. Moreover, we find that the SSIM metric does not always match the results of the APr preprocessing method, indicating that structural similarity might not adequately convey the perceptual quality enhancements brought about by adaptive range-bin selection. This motivates a need for better tailored metrics to specifically evaluate the quality of spectrograms and also align better with the model's perception when training a deep-learning-based denoiser. A good spectrogram quality metric should take into consideration important characteristics like background noise suppression and the clarity and continuity of ridge patterns that represent mD signatures.

%% file: tex/proposed_method.tex
The aforementioned techniques and assessments have concentrated on dynamic human activities that are distinguished by unique mD signatures; however, many real-world situations, particularly in settings like LTC facilities, also call for accurate identification of static or slowly changing activities (e.g., sitting, standing, watching television, transitioning from sitting to standing, etc.). The complete range of human behaviours that are present in these kinds of environments cannot be captured by dynamic activity recognition alone. Due to the lack of research in this field, it is especially intriguing. Static or slow-moving activities generate minimal Doppler shifts, rendering conventional mD spectrograms ineffective. As a result, the Doppler-time frequency spectrogram in section \ref{sec:dynamic_activities_evaluation} is not suitable for stationary objects, as it shows horizontal lines in the Doppler–time spectrogram, which is indistinguishable from static clutter and background noise. FMCW radar retains the ability to localize static reflectors through beat‑frequency processing and spatial filtering. By analyzing the phase difference between transmitted and received chirps, FMCW systems extract high‑resolution range profiles and, when combined with adaptive beamforming, we can obtain RA maps that can discriminate stationary objects in space \cite{infineon2024detection}.  Ali et al. \cite{fard2025exploring} leverage this capability by computing Capon RA maps and extracting spatial features to train machine learning classifiers for static pose recognition\cite{capon_algo}.

\begin{figure}[h!]
\centerline{\includegraphics[width=\linewidth]{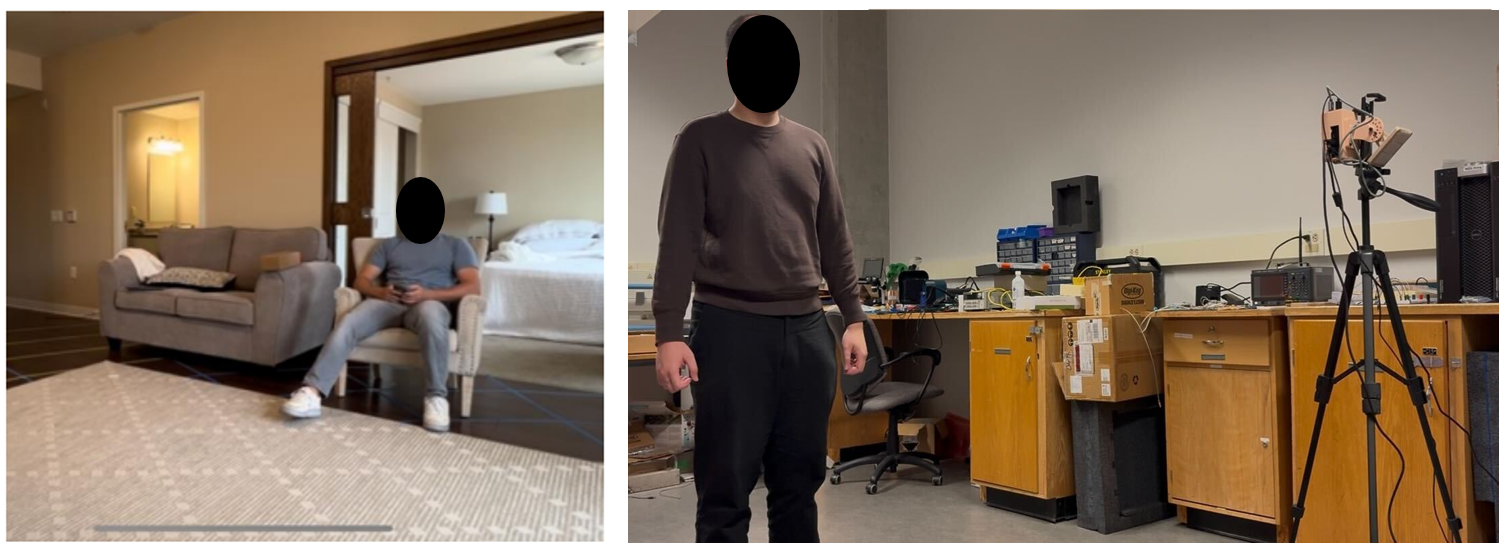}}
\caption{Our measurement setup in the long-term care facility (left) and our laboratory (right).}
\label{fig:measurement}
\end{figure}
To assess the suitability of existing RA‑map preprocessing for static activity recognition, we first reimplemented their pipeline comprising range‐FFT, range-Doppler spectrogram generation, and Capon beamforming and applied it to our collected dataset. For this study, we collected a dedicated dataset of static activities in two different environments (our laboratory as dataset $1$ and long‑term care facility as dataset $2$) using an off-the-shelf $60$\ GHz radar $BGT60TR13C$ radar from Infineon, with one transmitter and three receivers \cite{infineon2024bgt60tr13c}. Dataset 2 captures a range of basic postures: lying on floor (A0), lying on sofa (A1), sitting on floor (A2), sitting on sofa (A3), standing (A4), slow walking (A5) and minimal movements: getting up from sofa (A6), picking objects from floor (A7), and picking up objects from a table (A8) while dataset 1 was used for validation. The experimental setups are shown in the figure \ref{fig:measurement}. The resulting RA feature maps exhibit wide variability in spatial fidelity, as shown in figure \ref{fig:score_result}. To quantify how these map‐quality variations impact learning, we trained a lightweight 3D Convolution Neural Network (CNN) on the RA maps from dataset $2$. The model consists of three 3D convolutional blocks (kernel size of $3$, channel depths of $32,64,128$), each followed by batch normalization, ReLU activation, max-pooling and dropout, and $3$ fully connected layers. 
\begin{figure}[htbp]
\centerline{\includegraphics[width=\linewidth]{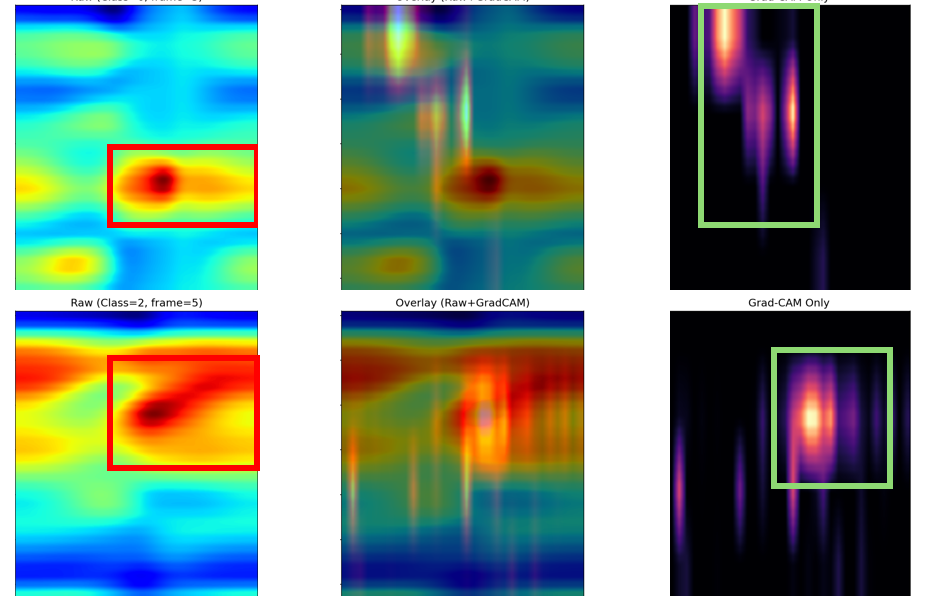}}
\caption{Grad‑CAM visualization on Range–Azimuth feature maps for frame $5$ of static activities. Top row: class A0 (lying on floor); bottom row: class A2 (sitting on sofa). In each row, the left panel shows the raw RA map with the red box indicating the ground‑truth target region, the center panel overlays the Grad‑CAM heatmap (magma colormap), and the right panel highlights the network’s attended region (green box).}
\label{fig:gradcam}
\end{figure}
We used Grad-CAM \cite{Selvaraju_2019} to visualize the network's class-discriminative attention on individual frames. Figure \ref{fig:gradcam} shows Grad‑CAM overlays for frame $5$ of class lying on floor (class $0$ - top row) and class sitting on floor (class $2$ - bottom row): the red box denotes the ground‐truth target blob, and the green box highlights the region to which the model attends. In the top row, the model incorrectly focuses on a background lump, whereas in the bottom example, its attention aligns well with the annotated target. These failure modes demonstrate that the CNN model can learn spurious clutter features when RA‐map quality is poor, leading to brittle static‐activity classification. Motivated by these observations, we propose no‑reference Range-Angle feature‐map 'Quality Scoring and 'Temporal Tracking' algorithms.

\subsection{Quality Scoring Algorithm}
To quantify the fidelity of the RA‐map without requiring a clean reference, we introduce a “lumps+subpeak” algorithm that produces a frame‐level \emph{cleanliness score} \(S\in[0,1]\) by combining three feature-map characteristics: (i) energy concentration in high‐intensity blobs, (ii) purity of each blob via sub‐peaks, and (iii) global peak prominence. 

\begin{algorithm}[ht]
\caption{Calculate frame score based on Lump + Peak Scores}
\label{alg:cleanliness}
\begin{algorithmic}[1]
% \REQUIRE 
%   A 2D radar frame $X \in \mathbb{R}^{H\times W}$, 
%   threshold percentile $\mathit{pct}$ (e.g.~88\%), 
%   penalty coefficient $\alpha$, 
%   minimum relative energy $\beta$ (e.g.~0.15), 
%   local-max distance $d_{\min}$ (e.g.~3), 
%   sub-peak threshold $\gamma$ (e.g.~0.9), 
%   peak-factor $\phi$ (e.g.~0.6).

% \ENSURE Cleanliness score $S \in [0,1]$.

\vspace{1mm}
\STATE $\mathit{lumps} \leftarrow \textsc{LumpsFromFrame}(X,\ \mathit{pct})$ 
  \COMMENT{find connected blobs above percentile}
% \IF {$\mathit{lumps}$ is empty}
%   \RETURN $0.0$
% \ENDIF

% \STATE Sort $\mathit{lumps}$ by descending sum of intensities ($\textit{sum\_i}$)
% \STATE $E_{\text{main}} \leftarrow \mathit{lumps}[0].\textit{sum\_i}$   \COMMENT{largest lump energy}
% \STATE $\mathit{validLumps} \leftarrow \{\}$

% \FORALL {lump $L$ in $\mathit{lumps}$}
%    \IF {$L.\textit{sum\_i} \ \geq\ \beta \times E_{\text{main}}$}
%        \STATE Append $L$ to $\mathit{validLumps}$
%    \ENDIF
% \ENDFOR

% \IF {$\mathit{validLumps}$ is empty}
%   \RETURN $0.1$
% \ENDIF
% \STATE $E_{\text{main}} \leftarrow \mathit{validLumps}[0].\textit{sum\_i}$ 
%   \COMMENT{updated largest among valid lumps}
% \STATE $E_{\text{total}} \leftarrow \sum_{L \in \mathit{validLumps}} L.\textit{sum\_i}$
% \STATE $\textit{leftover} \leftarrow E_{\text{total}} - E_{\text{main}}$
% \STATE $\mathit{leftoverRatio} \leftarrow \textit{leftover} / E_{\text{main}}$

\vspace{2mm}
\COMMENT{\textbf{detect sub-peak inside lumps}}
% ========== Sub-Peak detection inside lumps
\STATE $\textit{subpeakCount} \leftarrow 0$
\STATE $\textit{maxVal} \leftarrow \max(\textit{subArray})$
\STATE $\textit{absThresh} \leftarrow \gamma \times \textit{maxVal}$ 

\FORALL {lump $L$ in $\mathit{validLumps}$}
   % \STATE $\textit{bbox} \leftarrow L.\textit{bbox}$  \COMMENT{$(\text{min}_r,\text{min}_c,\text{max}_r,\text{max}_c)$}
   % \STATE $\textit{subArray} \leftarrow X[\textit{bbox}]$ \COMMENT{crop frame to this lump}
   \STATE $\mathit{coords} \leftarrow \textsc{PeakLocalMax}(\textit{subArray}, d_{\min}, \textit{absThresh})$
   \STATE $\textit{subpeakCount} \leftarrow \textit{subpeakCount} + \bigl|\mathit{coords}\bigr|$
\ENDFOR

\vspace{1mm}
\STATE $\textit{lumpsPenalty} \leftarrow \alpha \times (\textit{subpeakCount} - 1)$
\STATE $\textit{leftoverRatio} \leftarrow \dfrac{E_{\text{total}} - E_{\text{main}}}{E_{\text{main}}}$
\STATE $\textit{lumpsScore} \leftarrow \dfrac{1}{1 + \mathit{leftoverRatio} \;+\;\textit{lumpsPenalty}}$
\STATE $\textit{lumpsScore} \leftarrow \max\bigl(0,\ \min(\textit{lumpsScore}, 1)\bigr)$

\vspace{1mm}
% ========== Peakiness measure
\STATE $p \leftarrow \max(X)$ \COMMENT{global peak}
\STATE $\mu \leftarrow \mathit{mean}(X)$
\STATE $\sigma \leftarrow \mathit{std}(X) + 10^{-9}$
\STATE $\textit{peakness} \leftarrow \max\{0,\ (p - \mu)/\sigma\}$
\STATE $\textit{peakScore} \leftarrow 1 - \exp(-\phi \times \textit{peakness})$
\STATE $\textit{peakScore} \leftarrow \max\bigl(0,\ \min(\textit{peakScore}, 1)\bigr)$

\vspace{1mm}
% ========== Combine final
\STATE $\textit{w1} \leftarrow 0.2,\quad \textit{w2} \leftarrow 0.8$ 
   \COMMENT{weights}
\STATE $S \leftarrow (\textit{lumpsScore})^{\textit{w1}} + (\textit{peakScore})^{\textit{w2}}$
\STATE $S \leftarrow \max\bigl(0,\ \min(S, 1)\bigr)$
\RETURN $S$

\end{algorithmic}
\end{algorithm}

% \vspace{1em}

\begin{figure}[htbp]
\centerline{\includegraphics[width=\linewidth]{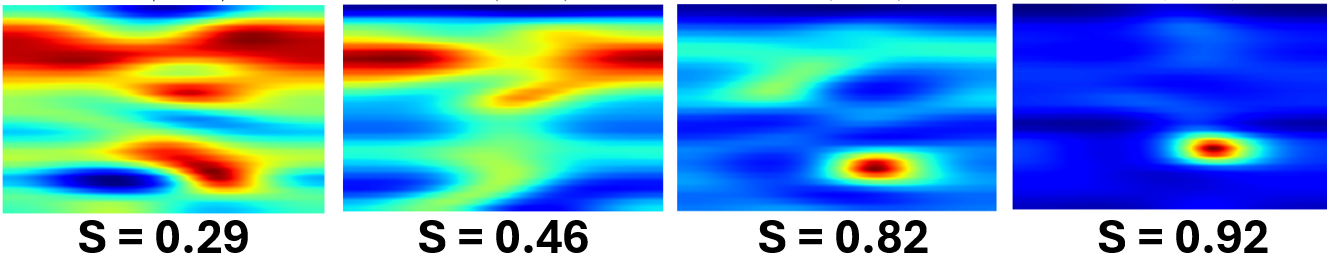}}
\caption{Scores calculated from our proposed algorithm corresponding to features map.}
\label{fig:score_result}
\end{figure}

High-intensity lump is often but not necessarily true scatter, while multiple scatters and subpeaks will result in penalty. These will capture cases of bad-quality feature maps that contain different lumps or a true lump that is a burden in the background. By empirical study, the hyperparameters were set to a percentile threshold of $\mathit{pct} = 90$, a subpeak penalty $\mathit{\alpha = 3}$ for single peak (and $\mathit{\alpha = 0.1}$ when multiple peaks are present), and fusion weight $\mathit{w_1} = 0.25$ and $\mathit{w_2} = 0.75$. These values were tuned according to dataset $2$ and then validated on dataset $1$, demonstrating that the cleanliness scorer generalizes robustly across two very different environments. The resulting scores obtained from algorithm \ref{alg:cleanliness} for representative RA maps are shown in Figure \ref{fig:score_result}.
% \textcolor{red}{Explain more neatly here how we come up with these formula, based on image analysis, "blobs-alike detection, energy concentration} 
% Let \(X\in\mathbb{R}^{H\times W}\) be one RA map. We find high-intensity lumps by thresholding $X$ at the given percentile $\mathit{pct}$, then labeling connected regions.  
% Let $E_{\text{main}}$ be the maximum intensity sum among valid lumps, and let $E_{\text{total}}$ be the sum over all lumps that exceed $\beta\,E_{\text{main}}$. 
% The \emph{leftover ratio} is 
% \begin{equation}
   % \mathrm{LR} \;=\; \frac{E_{\text{total}} - E_{\text{main}}}{E_{\text{main}}}.
% \end{equation}
% Next, we detect sub-peaks inside each lump.  Let $C$ = total number of local maxima above $\gamma \times \mathrm{maxVal}$ in each lump’s subarray.  The penalty grows with extra sub-peaks and lump \emph{lumps score} $\textit{LS}$ is
% \begin{equation}
%    \textit{lumpsPenalty} \;=\; \alpha \,(C - 1).
% \end{equation}
% \begin{equation}
%    \textit{LS} \;=\; \frac{1}{\,1 + \mathrm{LR} + \textit{lumpsPenalty}\,},
% \end{equation}
% clamped to $[0,1]$. Finally, we estimate \emph{peakiness} by comparing the global maximum of $X$ against the mean $\mu$ and standard deviation $\sigma$.  Define
% \[
%    \textit{peakness} \;=\; \max\bigl\{0,\; (p - \mu)\,/\,\sigma \bigr\},
% \]
% and
% \[
%    \textit{peakScore} \;=\; 1 \;-\; e^{-\phi\cdot \textit{peakness}}.
% \]
% The final cleanliness score $S$ is a geometric combination:
% \begin{equation}
%    S \;=\; (\textit{LS})^{w_1}\,\times\,(\textit{peakScore})^{w_2} \;\in\; [0,1].
% \end{equation}

% \subsection{Visual Explainability}

\subsection{Temporal Tracking Algorithm}
\begin{figure}[htbp]
\centerline{\includegraphics[width=\linewidth]{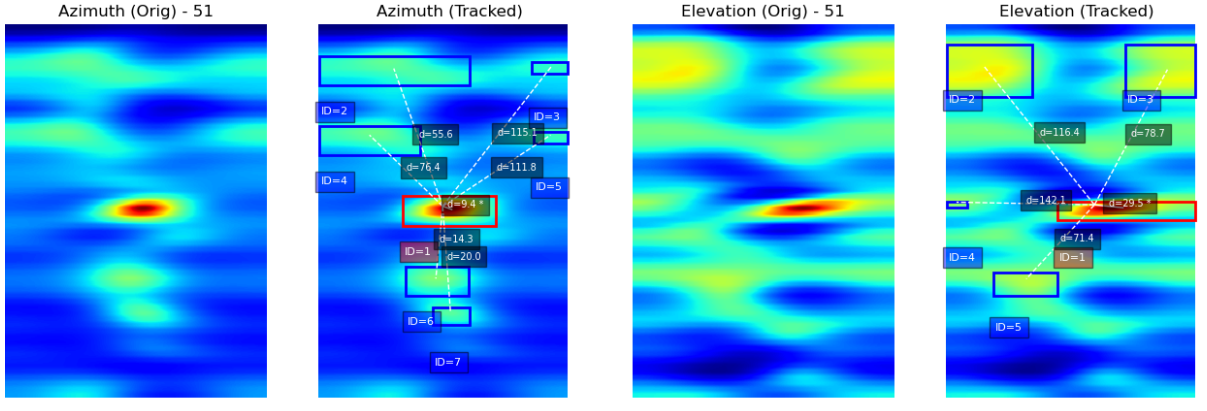}}
\caption{Range-Azimuth and Range-Elevation feature maps of frame 51 (class 2) and their tracking results (columns 2 and 4).}
\label{fig:frame51_tracking_algorithm}
\end{figure}

To enforce temporal consistency and generate reliable region proposals, we implement a lump‐tracking algorithm that links the primary scatterer across all frames.  At each frame \(t\), we first detect all high‐intensity lumps and compute their cleanliness scores \(S_t\) via Algorithm \ref{alg:cleanliness}.  If the feature map's score satisfies \(S_t\ge\tau\) threshold (we use \(\tau=0.8\)), we accept it immediately as the tracked target. Otherwise, we perform nearest‐centroid gating: we compare the previous frame’s centroid \(\mathbf{c}_{t-1}\) to each candidate lump \(\mathbf{c}_i\) and select the one with maximum energy within a distance threshold \(d\) (in our case \(d = 30\)). We maintain this matching over a sliding window of five frames that corresponds to $0.5$ seconds data for smooth transient dropouts. In Figure \ref{fig:frame51_tracking_algorithm}, the chosen lump is outlined in red, while all other detected lumps are shown in blue.

\begin{figure}[htbp]
\centerline{\includegraphics[width=\linewidth]{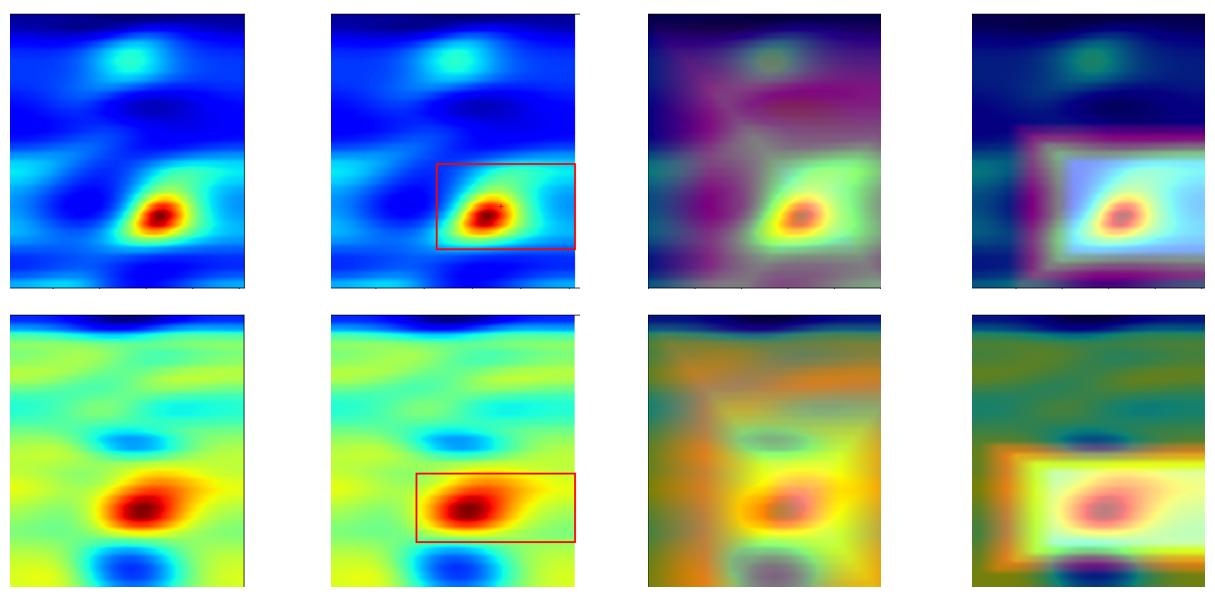}}
\caption{Example of Range-Azimuth maps and their computed masks. Columns 1-2: Raw RA maps and detected bounding box overlay. Column 3: Soft masks. Column 4: Hard masks.}
\label{fig:custom_mask}
\end{figure}

Once the target blob is identified, we generate $2$ complementary masks for downstream processing (Figure \ref{fig:custom_mask}): soft and hard masks. For the soft mask, weights radially decay from the tracked centroid to the whole maps, while weights are binary region of interest (ROI) (strictly $1$ within the bounding region and $0$ outside). These masks are then applied multiplicatively to each RA map, biasing the CNN’s attention toward the temporally consistent scatterer and suppressing background clutter. Together, our no‑reference quality scorer and temporal tracker will provide both precise region annotations for training and robust spatial priors at inferencing. This combination helps improve robustness and interpretability in radar‑based static activity recognition.

%% file: tex/conclusion.tex
Our project investigated radar-based HAR in both dynamic and static activity regimes. We used three recent preprocessing and denoising techniques for dynamic activities and evaluated them using both perceptual image quality metric SSIM and conventional error-based metrics, e.g., MSE, PSNR, etc. The findings demonstrate the shortcomings of traditional metrics in noisy settings and emphasize the necessity of perceptually aligned metrics for deep learning-based denoiser training. Our novelty features a radar spectrogram processing pipeline based on range-angle feature maps to tackle the frequently disregarded problem of static activity recognition. Among our contributions are a temporal tracking algorithm for consistency across frames and a no-reference “lumps+subpeak” quality scoring algorithm for measuring RA-map fidelity. By using these techniques, clutter is reduced and the model's focus is directed toward significant spatial features.

Our study suggests, radar-based HAR systems' resilience is greatly increased when adaptive signal processing and customized spatial analysis are combined. Future research will integrate our tracking-and-masking framework into end-to-end deep learning pipelines for continuous activity monitoring in real-world scenarios and investigate joint time-frequency-spatial resolution adaptation.

% \textcolor{red}{Wrap up, summarize each contribution since there are few of them, and provide brief future research direction}